\documentclass[conference]{IEEEtran}
\IEEEoverridecommandlockouts
\usepackage{cite}
\usepackage{amsmath,amssymb,amsfonts}
\usepackage{mathtools}
\usepackage{braket}  
\usepackage{bm}    
\usepackage{graphicx}
\usepackage{textcomp}
\usepackage{xcolor}
\usepackage{url}
\usepackage{tabularx}
\usepackage{subfig}
\usepackage{comment}
\usepackage{multirow}
\def\tablename{Table}
\usepackage{etoolbox}
\usepackage{tikz}
\usetikzlibrary{shapes.geometric, arrows.meta, positioning}
\usepackage{algorithm}
\usepackage{algpseudocode}
\algrenewcommand\algorithmicrequire{\textbf{Input:}}
\algrenewcommand\algorithmicensure{\textbf{Output:}}
\makeatletter
\patchcmd{\@makecaption}
  {\scshape}
  {}
  {}
  {}
\makeatletter
\patchcmd{\@makecaption}
  {\\}
  {.\ }
  {}
  {}
\makeatother
\def\tablename{Table}
\def\BibTeX{{\rm B\kern-.05em{\sc i\kern-.025em b}\kern-.08em
    T\kern-.1667em\lower.7ex\hbox{E}\kern-.125emX}}
\begin{document}

% The mapping of quantum algorithms in multi-core quantum computing architectures

\title{Efficiently architecting VQAs: Expressibility--Trainability--Resources Pareto-Optimality\\
}

\author{\IEEEauthorblockN{Rodrigo M. Sanz\textsuperscript{1}, Andreu Anglés-Castillo\textsuperscript{1}, Eduard Alarcón\textsuperscript{2}, Carmen G Almudéver\textsuperscript{1}}
\IEEEauthorblockA{\textsuperscript{1}\textit{Universitat Politècnica de València, Valencia, Spain}}
\IEEEauthorblockA{\textsuperscript{2}\textit{NaNoNetworking Center in Catalonia (N3Cat), Universitat Politecnica de Catalunya, Barcelona, Spain}}}

\maketitle

\begin{abstract}
Ansatz selection is a key factor in the performance of variational quantum algorithms (VQAs). While much of the state-of-the-art still relies on heuristic choices, an inadequate circuit structure can compromise both the expressive power and the trainability of the resulting model. Recent results have also established theoretical connections between expressibility and the onset of barren plateaus, highlighting the need for systematic criteria for ansatz selection.
In this work, the ansatz is treated as a design feature to be optimized rather than a fixed block, and a design space exploration (DSE) is performed over a diverse set of parametrized quantum circuits (PQCs). Three complementary metrics---expressibility, trainability, and resource cost---are evaluated and used to analyze the trade-offs that emerge across different PQCs. Beyond identifying Pareto-optimal candidates, this multi-objective perspective helps clarify the interplay between these metrics and contributes quantitative evidence toward understanding the expressibility--trainability tension in variational circuits.
\end{abstract}


%%% Sections %%%
\section{Introduction}

Quantum computing (QC) has emerged as a promising paradigm for addressing computational tasks that are intractable for classical computers. In particular, quantum algorithms have been shown to provide rigorous complexity-theoretic advantages over their classical counterparts in problems such as integer factorization \cite{Shor1997}, quantum simulation \cite{Lloyd1996, LowChuang2019}, and resolution of linear systems of equations \cite{HHL2009}, among others. This potential has driven intense efforts toward the development of scalable quantum hardware \cite{yoder2025tourgrossmodularquantum}.
However, none of these algorithms have yet been executed on real quantum hardware in a regime that demonstrates a clear advantage over the best known classical methods, as they typically require deep circuits and fault-tolerant devices with a large number of qubits. This motivates the need for strategies that enable efficient implementations under realistic hardware constraints, explicitly accounting for resource metrics such as circuit depth or entangling-gate count, while also incorporating device-level limitations such as restricted connectivity and limited gate set. Accordingly, as we transition from the NISQ (Noisy Intermediate-Scale Quantum, \cite{Preskill2018NISQ}) to the FASQ (Fault-tolerant Application-Scale Quantum, \cite{Preskill2025FASQ}) era, resource-aware optimization becomes essential to make quantum algorithms executable, scalable, and ultimately verifiable on hardware.

\begin{figure*}[t]
    \centering
    \includegraphics[width=1\textwidth]{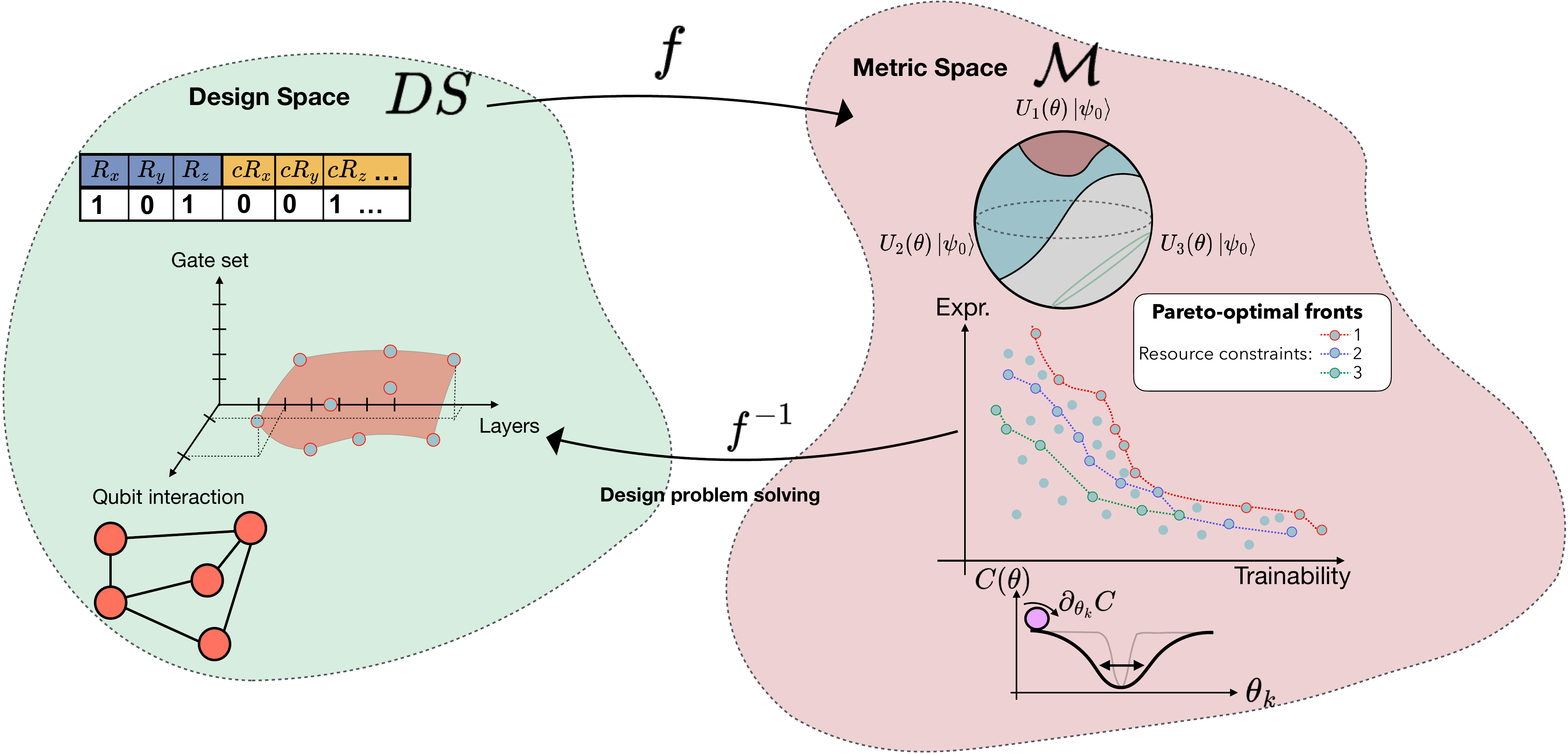}
    \caption{Design-space to metric-space mapping for resource-aware ansatz search.
\textbf{Left:} schematic view of the design space $DS$, whose axes encode key architectural choices for a PQC family: the primitive gate set, qubit interaction-graph, and number of layers.
\textbf{Right:} the induced metric space $\mathcal{M}$, where each circuit $U(\boldsymbol{\theta})\ket{\psi_0}$ is represented by performance descriptors, namely expressibility, trainability, and a resource-cost metric, revealing the expressibility--trainability trade-off and the corresponding Pareto-optimal fronts under different resource constraints.
Overall, the workflow consists of (i) mapping candidate PQCs from $DS$ to $\mathcal{M}$ via $f$ to identify Pareto-optimal circuits, and (ii) using $f^{-1}$ to project the selected family back onto $DS$ to localize the region of interest within the design space.}
    \label{fig:triangle}
\end{figure*}

In this regime, a representative class of NISQ-oriented methods comprises Variational Quantum Algorithms (VQAs) \cite{Cerezo2021VQAReview}. In contrast to many quantum algorithms originally conceived for fault-tolerant machines, VQAs leverage shallow circuit primitives, which helps mitigate the impact of noise and hardware limitations. Inspired by the empirical success of classical neural networks (NNs), VQAs rely on parametrized quantum circuits (PQCs), or ansätze, whose parameters (rotation angles of quantum gates) are optimized through a classical feedback loop to minimize a problem-dependent cost function. This hybrid quantum--classical paradigm has enabled a wide range of applications, including machine learning tasks \cite{Havl_ek_2019, Schuld_2019,BravoPrieto2023variationalquantum}, combinatorial optimization \cite{farhi2014quantumapproximateoptimizationalgorithm,PhysRevX.10.021067}, and ground-state preparation \cite{Peruzzo_2014,Kandala_2017}.

Despite their versatility and practical relevance, the performance of VQAs is strongly influenced by the structural properties of the chosen ansatz. In particular, it is well understood that fundamental trade-offs exist between the expressibility of a circuit, its implementation cost, and its trainability, i.e., how efficiently gradients can be estimated during optimization. Highly expressive circuits are often necessary to represent the solution to complex problems, yet they can incur large hardware costs or suffer from optimization difficulties, such as vanishing gradients in certain regimes \cite{Larocca2025BarrenPlateaus}.

To date, VQAs are largely employed in a heuristic manner, relying on ansätze that have been empirically shown to converge for a given task. This practice provides partial insight into whether a given PQC is well matched to the problem constraints or whether a different circuit could still show a good enough performance in terms of accuracy and trainability while requiring a moderate amount of resources (e.g. low circuit depth). This naturally motivates a more rigorous strategy: systematically exploring the Design Space (DS) of PQCs with the aim of finding circuit designs that achieve an optimal compromise among expressibility, trainability, and implementation cost, as schematically illustrated in Fig.~\ref{fig:triangle}. This implies treating the ansatz as a design choice to be optimized under competing objectives instead of a fixed circuit block.

Following this rationale, in this work, a design space exploration (DSE) \cite{GautierACK2022Sherlock, NardiKO2019PracticalDSE} methodology is adopted to study PQCs in a systematic and task-aware manner. For a representative problem instance with global correlations (Transverse-Field Ising Model, TFIM), state-of-the-art metrics used to quantify circuit expressibility, trainability, and resource requirements are evaluated, and the trade-offs that emerge across a diverse family of ansätze are analyzed. This approach enables the identification of Pareto-optimal circuits with respect to the objective metrics simultaneously. Beyond providing a framework for ansatz selection and design, the results offer empirical evidence of the expressibility--trainability relationship, whose exact mathematical dependence remains an open question\footnote{Although no exact analytic relationship was found between gradient magnitudes and ansatz expressibility, relevant results derived in \cite{Holmes2022ExpressibilityGradients} set lower bounds on such scalings.} and is key to assessing whether variational quantum computing can scale feasibly to larger problem sizes.

Overall, the methodology proposed in this work provides a practical route for resource-aware circuit design under realistic hardware constraints. Although we focus on VQAs as a representative case study, the same framework and methodology can be readily applied to a broader range of applications, including quantum compilation benchmarking, error mitigation and noise-aware circuit design, pulse-level gate control, and quantum error correction, among others.

\newpage
\section{Background and prior work}\label{Sec: II}

\subsection{Variational Quantum Algorithms}\label{Sec: II.a}

As introduced above, Variational Quantum Algorithms constitute a general hybrid quantum--classical framework to tackle a broad range of tasks on near-term quantum hardware. Despite presenting different variations depending on the particular application, their core structure is shared across settings: a parametrized quantum circuit or \emph{ansatz} generates a quantum state given a set of parameters $\ket{\psi_{\boldsymbol{\theta}}}=U(\boldsymbol{\theta})\ket{0}^{\otimes n}$, a problem-dependent \emph{cost function} assigns a scalar figure of merit $C(\boldsymbol{\theta})$, and a \emph{classical optimizer} updates the parameters to minimize such function. The variational problem can therefore be written as finding parameters $\boldsymbol{\theta}^\star$ such that
\begin{equation}
    \boldsymbol{\theta}^\star \in \arg\min_{\boldsymbol{\theta}}\, C(\boldsymbol{\theta}) .
\end{equation}

On the one hand, the ansatz is typically expressed as a composition of parameterized blocks,
\begin{equation}
    U(\boldsymbol{\theta}) = U_L(\boldsymbol{\theta}_L)\cdots U_2(\boldsymbol{\theta}_2)\,U_1(\boldsymbol{\theta}_1),
\end{equation}
where each $U_\ell(\boldsymbol{\theta}_\ell)$ generally consists of elementary gates drawn from a chosen primitive gate set and connectivity pattern. On the other hand, the cost function is most commonly defined as the expectation value of a quantum observable $O$, whose choice encodes the target task (e.g. an objective Hamiltonian in ground-state preparation).
\begin{equation}
    C(\boldsymbol{\theta}) = \langle O \rangle_{\boldsymbol{\theta}}
    = \bra{\psi_{\boldsymbol{\theta}}} O \ket{\psi_{\boldsymbol{\theta}}}.
\end{equation}
In practice, this quantity is not directly accessed but estimated from measurement outcomes obtained by repeatedly sampling the same prepared state $\ket{\psi_{\boldsymbol{\theta}}}$ so as to obtain a statistically reliable estimate of $C(\boldsymbol{\theta})$ at each optimization step. The estimated cost is then returned to the classical routine, closing a feedback loop that is iteratively updated $\boldsymbol{\theta}$ until convergence.

Crucially, the ansatz specifies the variational family of accessible states, thereby defining a variational manifold $\mathcal{V}(\boldsymbol{\theta})$ \cite{Hackl2020GeometryVariationalMethods}. As a result, it can determine whether a solution to the target problem is representable by the algorithm at all, independently of the optimizer, as well as how difficult it is to reach such a solution through training. This makes ansatz design a central factor in the performance of VQAs and motivates the need for principled criteria to compare circuit structures. The subsequent sections introduce the three complementary metrics used throughout this work, which together characterize, respectively, the coverage of the Hilbert space, the optimization landscape induced by a chosen objective, and the resources required to implement a particular PQC.

\subsection{Expressibility of a Parametrized Quantum Circuit}\label{Sec: II.b}

The expressibility of a PQC can be understood as its capability to \textit{uniformly sample} from the Hilbert space. In \cite{Sim2019Expressibility}, it is shown that this can be quantified by comparing its deviation from the Haar measure (i.e., the maximally expressive scenario) \cite{Mele_2024}. In particular, this is done via the probability density function (PDF) of state overlaps:

\begin{equation}
    P(F=|\braket{\psi_{\bm{\theta}}|\psi_{\bm{\theta}'}}|^2)
\end{equation}
where $\ket{\psi_{\bm{\theta}}}$ and $\ket{\psi_{\bm{\theta}'}}$ are states sampled from the distribution that we aim to characterize, and $\bm{\theta} \coloneqq (\theta_1, ..., \theta_n)$ ,  $\bm{\theta}' \coloneqq (\theta_1', ..., \theta_n')$ are arrays of uniformly random parameters in the continuous interval $[0,2\pi)$. 

\begin{figure}[h]
    \centering
    \includegraphics[width=1\linewidth, trim=0 100 0 100, clip]{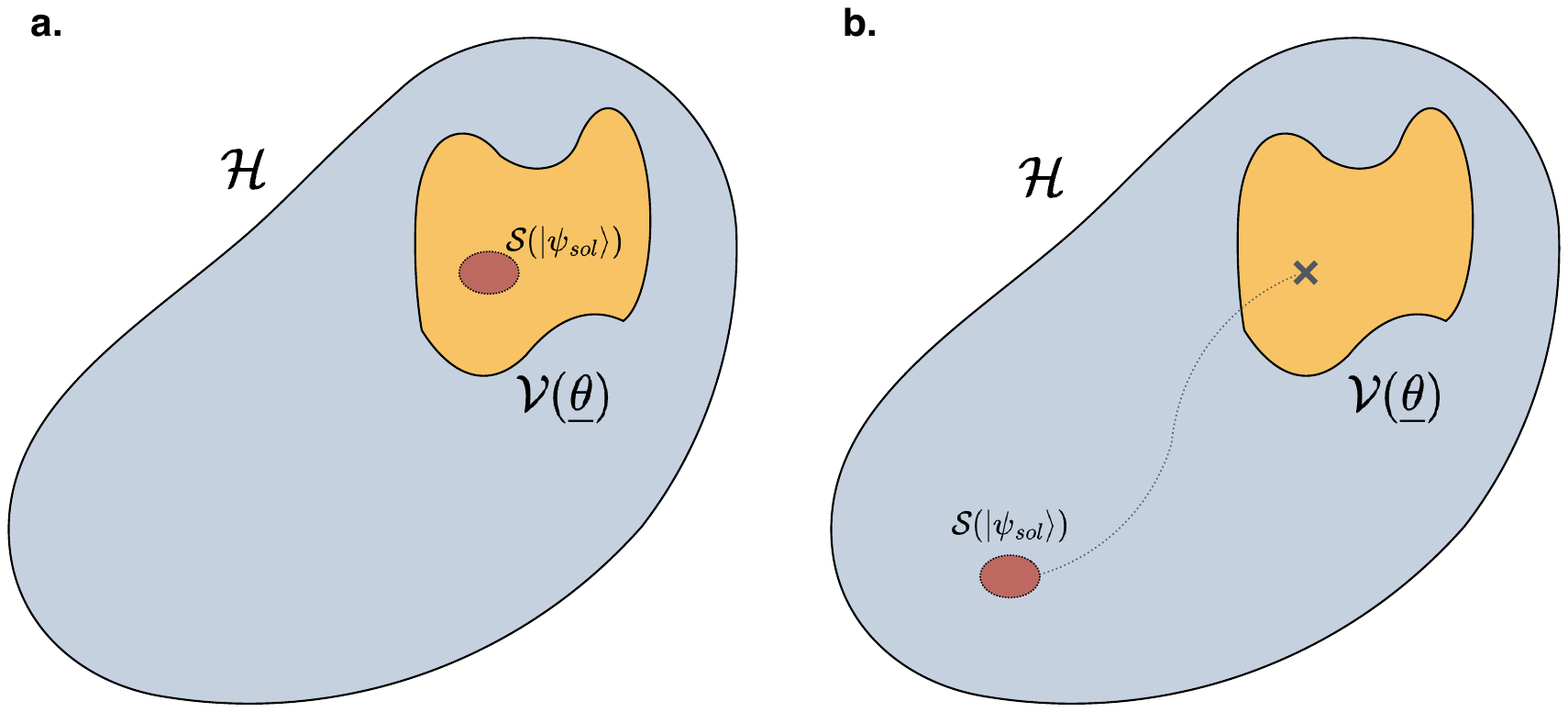}
    \caption{ Visual representation of the Hilbert space (blue), the set of states reachable by the ansatz (yellow), and the solution subspace (red). In (a), the PQC is \emph{complete} for the given task, whereas in (b) it is \emph{incomplete}, i.e., the solution is not contained within the ansatz domain.}
    \label{fig:expr_hilbert}
\end{figure}

Even though the full form of $P(F)$ for a given ansatz is in general unknown, it can be statistically estimated by independently sampling pairs of states from it, computing its overlap and counting the results in a histogram. This strategy provides a discrete version of the aimed distribution, which, of course, will suffer from a finite sampling error. Let us denote such an estimate as $\tilde{P}_{PQC}(F)$. Nonetheless, the analytic expression of the state fidelity PDF for the Haar distribution in the $N$-dimensional Hilbert space is well known: $P_{Haar}(F) = (N-1)(1-F)^{N-2}$ \cite{Zyczkowski2005AverageFidelity}. Hence, the difference between $\tilde{P}_{PQC}(F)$ and a discretized version of $P_{Haar}(F)$ can be used as a proxy to quantify the expressibility of a PQC (i.e, the deviation from the Haar measure). In particular, in \cite{Sim2019Expressibility} it is proposed to use the Kullback-Leibler (KL) divergence \cite{Kullback1951OnIA}:

\begin{equation}\label{eq:Expr}
    \text{Expr} = D_{KL}(\tilde{P}_{PQC}(F)||P_{Haar}(F))
\end{equation}
This quantity, therefore, informs about the size and sparsity of the variational manifold \cite{Hackl2020GeometryVariationalMethods} induced by the ansatz. From this perspective, it is natural to relate expressibility to the notion of \emph{completeness} of an ansatz for a given task \cite{Holmes2022ExpressibilityGradients}. An ansatz is \emph{complete}  (scenario illustrated in Fig.\ref{fig:expr_hilbert}a) for a given problem if the solution of such a task is contained within the family of reachable states generated by the PQC. Thus, given an arbitrary problem, the more expressive the ansatz is, the more probability it has to be \emph{complete}.  A simple example would be to consider a fully expressive PQC that spans the whole Hilbert space, that would be complete for any problem, as all solution states (or subspaces) must lie within it. On the other hand, if the ansatz is \emph{incomplete} (scenario illustrated in Fig.\ref{fig:expr_hilbert}b), then the maximum fidelity achievable by our model would be $F_{\text{max}} = |\braket{\psi^*|\psi_{\text{sol}}}|^2$, where $\ket{\psi^*}$ stands for the \emph{closest} state to the target within the PQC manifold\footnote{Since the solution space can be degenerated, $F_{\text{max}}$ will be in general different for each basis element.}. Note that, in this case,  even a successful training without any Barren Plateaus or local minima would provide an error $\epsilon = 1-F_{\text{max}}$.

Therefore, it is reasonable to expect that more expressive PQCs will generally yield better performance in VQAs. Indeed, this was explicitly tested in \cite{Hubregtsen2021EvaluationPQC}, where a set of ansätze was evaluated across nine datasets of increasing difficulty, obtaining an average Pearson Product-Moment Correlation Coefficient \cite{Ly2018AnalyticPosteriorsPearson} between expressibility and classification accuracy of $0.7\pm 0.05$, indicating a moderate-strong positive correlation.

As discussed in the following sections, increasing expressibility can compromise trainability and usually requires deeper circuits, especially in high–qubit-count regimes. Therefore, the impact of these correlation results on the broader ansatz-design challenge is limited, since they were observed only for models with fixed dimensionality. Consequently, an optimal ansatz choice is not achieved simply by maximizing expressibility in isolation, but rather by balancing the inherent trade-offs between expressibility, trainability, and implementation cost.

\subsection{Barren Plateaus}\label{Sec: II.c}

It is well known that Variational Quantum Algorithms can suffer from what is often called \emph{vanishing gradient phenomenon} as the number of qubits increases. This was first stated in \cite{McClean2018BarrenPlateaus}, where the gradient variance was shown to decrease exponentially with the number of qubits $n$ for PQCs that are (approximately\footnote{Here ``approximately'' means that the ansatz can be partitioned into two blocks \(U^+\) and \(U^-\), where at least one of them forms an exact unitary 2-design.}) 2-designs, i.e., circuits that match the Haar measure statistics up to the second moment $t=2$ \cite{Dankert_2009}.

This result was later extended to arbitrary PQCs by Holmes et al. in \cite{Holmes2022ExpressibilityGradients}, where the variance of the gradient of the ansatz is upper-bounded by the exponentially small factor found in \cite{McClean2018BarrenPlateaus} plus a term that depends on the \emph{expressibility} of the circuit. Notably, the second term vanishes if the circuit is a 2-design, recovering the original result from McClean et al. Even though this analytical result is a bound, it points towards a clear trade-off between \emph{expressibility} and \emph{trainability} in VQAs.

One might wonder why the variance of the gradient matters. This can be understood from the fact that one can prove that the expected value of the derivative of the cost function with respect to \textit{any} trainable parameter $\theta_k$ is zero \cite{McClean2018BarrenPlateaus}. In other words:

\begin{equation}\label{eq: unbiased}
    \left\langle \frac{\partial E}{\partial \theta_k} \right\rangle_{\bm{\theta}} = 0
\end{equation}
meaning that the cost landscape is \emph{unbiased}. Thus, one hopes to find with arbitrary high probability derivatives that are far enough from the average in order to efficiently train our models. Standard Chebyshev’s inequality bounds such a probability by a term that is proportional to the variance of the derivative distribution:

\begin{equation}
    P(|\partial_k E|\geq \delta) \leq \frac{\text{Var}[\partial_kE]}{\delta^2}
\end{equation}
Hence, the individual parameter gradient variances provide a practical indicator of PQC trainability, as they quantify how strongly gradients concentrate around zero the across the parameter space.

\subsection{Resources}\label{Sec: II.d}

Beyond expressibility and trainability, PQCs must ultimately be efficiently implementable on a target device. On NISQ hardware, the accumulated effect of gate and measurement errors and finite coherence times can dominate the output of the algorithms, to the point that improvements in circuit design become practically irrelevant if the resulting implementation is too noisy. Indeed, recent work \cite{Fontana2025ClassicalSimNoisyVQCs}--\cite{Angrisani2025PauliPropagationLocalNoise} shows that under certain assumptions it is possible to efficiently estimate expectation values coming from noisy quantum circuits classically. Moreover, it has also been shown that VQAs can present noise-induced Barren Plateaus \cite{Wang2021NoiseInducedBP}, implying that noise not only affects the accuracy of our protocol, but also fundamentally alters the optimization landscape.

From a practical standpoint, resource requirements must therefore be taken into account when selecting an ansatz. These requirements include circuit depth and gate counts (as proxies for noise accumulation), as well as the number of circuit evaluations required to estimate expectation values and gradients with sufficient statistical confidence. In this work, such considerations are explicitly incorporated through a dedicated \emph{resource} metric in the multi-objective optimization. 

Furthermore, such implementations must satisfy hardware constraints such as limited qubit connectivity and restricted native gate sets, among others. Satisfying these constraints typically requires a compilation (mapping) step, which can introduce routing overhead by increasing both the gate count and the circuit depth. In turn, given limited coherence times and gate-induced errors, this overhead generally reduces the overall circuit fidelity \cite{LiDX19,Lao_2022, Sivarajah_2020}. In order to account for this, the optimal PQCs are mapped into their design space (see Sec.\ref{sec:DSE}). This enables determining which circuit structures satisfy the hardware constraints, minimizing post-compilation overhead.

Taken together, these considerations motivate a methodology for ansatz selection that goes beyond the dominant practices in the literature. Indeed, most of the progress in VQAs has focused on better understanding the behavior of specific, well-known ansätze \cite{Nakaji2021ExpressibilityALA, Park2024hamiltonian, Leone2024practicalusefulness}, and in developing optimizations tailored to particular problem instances, such as quantum chemistry \cite{Tang_2021, Yordanov_2021} and combinatorial tasks \cite{zhu2022adaptivequantumapproximateoptimization,majumdar2021optimizingansatzdesignqaoa}. Yet, these advances do not provide a general resource-aware strategy for ansatz design/selection. In contrast, this work proposes to treat the ansatz as a design variable and introduces a systematic methodology to identify circuits that explicitly balance expressibility and trainability while satisfying practical resource constraints.

\section{Methodology}

In a Design Space Exploration \cite{NardiKO2019PracticalDSE, DSE_Carmina_Eduard}, a set of objects is represented in both a design space and a performance (metric) space. The design space is spanned by features that can be directly manipulated by the architect, whereas the metric space captures how each combination of design choices translates into performance under a given multi-objective optimization problem. After performing the optimization, one can identify regions of interest in the design space, which naturally enables the selection of optimal designs and the synthesis of new high-performing ones.

The goal of this work is to apply such a methodology to ansatz selection in VQAs, while gaining some understanding from the different trade offs between the objective functions of interest. To this end, the proposed metrics must be evaluated over a sufficiently representative family of parameterized quantum circuits. Such a family should span different application domains and encompass a variety of generators (primitive gate sets), connectivities, circuit depths, and numbers of trainable parameters. While many choices are possible, the present study focuses on the circuit family shown in Fig.~\ref{fig:appendix_circuits}, originally introduced by Sukin Sim et al.~\cite{Sim2019Expressibility}. More concretely, their corresponding versions with $L=1,2,3$ layers are considered for the final analysis. This latter choice of circuit depth was empirically found to be representative enough, because of saturation of the trainability and expressibility metrics. Indeed, in Sec.~\ref{sec: train_depth}, where more layered PQCs are evaluated for the trainability--cost study, a $\sim 4 \%$ average relative change in trainability was already found when increasing $L$ beyond 3.

For a fair comparison, the study is carried out at a fixed model dimensionality (namely, the same number of qubits for all PQCs). In order to illustrate the proposed protocol and to ensure consistency with the state of the art, the case $n=4$ qubits is selected as a reference example. However, the methodology naturally extends to arbitrary $n$, at the expense of increased computational cost in the evaluation of expressibility and trainability \cite{Sim2019Expressibility}. More details about scalability are addressed in the final discussion. 

Once the set of circuits to be characterized has been selected, the methodology proceeds with the definition of the metrics used to quantify the \emph{expressibility}, \emph{trainability}, and \emph{cost} of the PQCs, together with a detailed description of how each of these quantities is computed.

\subsection{Expressibility of the models}

As explained in Sec.~\ref{Sec: II.b}, one can quantify the expressibility of a PQC by the KL divergence between its pairwise fidelity distribution and the one from Haar-random sampled states. This can be numerically estimated by means of the following protocol:

\begin{algorithm}[h!]
\caption{Numerical estimation of PQC expressibility}
\label{alg:expressibility}
\begin{algorithmic}[1]
\Require Ansatz unitary $U(\boldsymbol{\theta})$, number of qubits $n$, number of parameters $|\mathcal{P}|$, number of pairs $N_{\text{pairs}}$, number of bins $n_{\text{bins}}$, Haar fidelity distribution $P_{\text{Haar}}(\mathcal{F})$
\Ensure Expressibility $\mathrm{Expr}'$
\State Initialize list $\mathcal{S} \gets \emptyset$ \Comment{store fidelities}
\For{$k = 1$ \textbf{to} $N_{\text{pairs}}$}
    \State $(\boldsymbol{\theta}_1,\boldsymbol{\theta}_2) \gets \mathrm{UniformRandom}([0,2\pi)^{2|\mathcal{P}|})$
    \State $\ket{\psi_{\boldsymbol{\theta}_1}} \gets U(\boldsymbol{\theta}_1)\ket{0}^{\otimes n}$ \Comment{State preparation}
    \State $\ket{\psi_{\boldsymbol{\theta}_2}} \gets U(\boldsymbol{\theta}_2)\ket{0}^{\otimes n}$
    \State $\mathcal{F}_k \gets \left|\braket{\psi_{\boldsymbol{\theta}_1}|{\psi_{\boldsymbol{\theta}_2}}}\right|^2$
    \State $\mathcal{S}$.append( $\mathcal{F}_k$ )
\EndFor

\State Build histogram $\hat{P}_{\text{PQC}}(\mathcal{F})$ from $\mathcal{S}$ using $n_{\text{bins}}$ bins
\State $\hat{P}_{\text{PQC}}(\mathcal{F})$ $\gets$ Histogram ($\mathcal{S}$ , $n_{\text{bins}}$ )

\State $D_{\mathrm{KL}} \gets D_{\mathrm{KL}}\!\left(\hat{P}_{\text{PQC}}(\mathcal{F}) \,\|\, P_{\text{Haar}}(\mathcal{F})\right)$
\State $\mathrm{Expr}' \gets -\log_{10}\!\left(D_{\mathrm{KL}}\right)$
\State \Return $\mathrm{Expr}'$
\end{algorithmic}
\end{algorithm}

%comment
\iffalse
\begin{itemize}
    \item Sample a pair of random parameter vectors $\bm{\theta}_1, \bm{\theta}_2$ where each element lies in the interval $[0,2\pi)$.

    \item Compute the pair of states $\ket{\psi_{\bm{\theta}_1}}$, $\ket{\psi_{\bm{\theta}_2}}$ by propagating an initial state (typically fixed to $\ket{0}^{\otimes n}$) through the ansatz as:
    \begin{equation}
        \ket{\psi_{\bm{\theta}}} = U(\bm{\theta})\ket{0}^{\otimes n}
    \end{equation}

    \item Then, compute the fidelity $\mathcal{F}_{\bm{\theta}_1,\bm{\theta}_2} = |\braket{\psi_{\bm{\theta}_1}|\psi_{\bm{\theta}_2}}|^2$.

    \item Repeat the process $N_{pairs}$ times. Then, we classify all fidelity values in $n_{bins}$ intervals and we retrieve the discrete distribution $\hat{P}_{PQC}(\mathcal{F})$.

    \item Finally, take the KL divergence with respect to the Haar fidelity distribution according to Eq.\ref{eq:Expr} to retrieve our estimate for our ansatz expressibility.
\end{itemize}
\fi
%end of comment

For the simulations, the number samples and bins in the histogram are set to $N_{pairs}=5000$ and $n_{bins}=75$ respectively, motivated by \cite{Sim2019Expressibility}, where these values were found to be enough for robust and stable convergence.

Moreover, in this study the expressibility is quantified as the negative logarithm of the KL divergence, as first proposed in \cite{Hubregtsen2021EvaluationPQC}. 

\begin{equation}\label{eq:expr'}
     \text{Expr'}= -\log_{10}{D_{KL}}
 \end{equation}
Importantly, according to this new reformulation, the metric gets larger as the circuit becomes more expressive, contrary to the original definition. This slight change is motivated mainly to better discriminate between highly-expressive circuits that may present similar KL divergences (arbitrarily close to 0) but can be easily (i.e, with less samples) comparable by means of Expr'. Consequently, it was found to correlate more with performance in applications such as classification tasks in \cite{Hubregtsen2021EvaluationPQC}. 

\subsection{Trainability}

In contrast to expressibility and cost-related metrics, which can be defined independently of the target problem, the notion of trainability is inherently tied to a specific cost function. As a consequence, any quantitative assessment of trainability must be carried out with respect to a concrete observable defining the optimization landscape.

In our analysis, trainability is evaluated by considering a physically motivated and widely studied problem, namely the Transverse-Field Ising Model (TFIM) with open boundary conditions. The corresponding Hamiltonian is given by

\begin{equation}
    H = -J \sum_{i=1}^{n-1} \sigma_i^z \sigma_{i+1}^z - h \sum_{i=1}^{n} \sigma_i^x ,
\end{equation}
where $J$ denotes the nearest-neighbor Ising coupling and $h$ the strength of the transverse field. For simplicity, these parameters are set to $J=h=1$ in all simulations.

Beyond its broad relevance in quantum many-body physics and quantum simulation, the TFIM is also a natural choice from a generality standpoint, as it provides a global cost function, involving operators supported over all qubits and multiple nearest-neighbor two-qubit correlation terms. To further explore the influence of cost-function structure on trainability, given the theoretical results linking locality to the emergence of barren plateaus \cite{Cerezo_2021}, the analysis is extended in Appendix \ref{apx:other_models}. There, the isotropic Heisenberg model and a purely local-X Hamiltonian (the $J=0$ limit of the TFIM) are analyzed. Collectively, these cases provide a comprehensive map of how various PQCs adapt to different landscape complexities and interaction symmetries.

The trainability of a given pair $\{U(\boldsymbol{\theta}), H\}$ is assessed through the gradient behavior of the associated cost function,
\(
E(\boldsymbol{\theta}) = \langle \psi_{\boldsymbol{\theta}} | H | \psi_{\boldsymbol{\theta}} \rangle ,
\)
with respect to the trainable parameters. As discussed in Sec.~\ref{Sec: II.c}, the scaling of gradient variances provides a direct diagnostic of the optimization landscape: large variances are associated with favorable trainability, whereas exponentially suppressed variances indicate the presence of barren plateaus. Building on these considerations, in this work an averaged sum of the gradient variance is selected,

\begin{equation}
    \text{Trainability} = \frac{1}{|\mathcal{P}|}\sum_{k\in \mathcal{P}} \mathrm{Var}\!\left[\partial_k E\right],
\end{equation}
where $\mathcal{P}$ denotes the set of all trainable parameters of the model.

The numerical procedure used to estimate this quantity is summarized in Algorithm~\ref{alg:grad_variance}, which outlines the steps required to sample the parameter space, evaluate the gradients\footnote{The function ''$\text{Gradient}\!\left(E\right)\big|_{\boldsymbol{\theta}^{(i)}}$'' in Algorithm~\ref{alg:grad_variance} inherently calls the well-known Parameter-Shift Rule subroutine \cite{Wierichs_2022} for gradient estimation.} of the TFIM cost function, and compute the resulting average gradient variance for each circuit.

\begin{algorithm}[h!]
\caption{Estimation of the average gradient variance}
\label{alg:grad_variance}
\begin{algorithmic}[1]
\Require Ansatz unitary $U(\boldsymbol{\theta})$, observable $E$, number of samples $N_s$, trainable parameters $\{\theta_k\}_{k=1}^{|\mathcal{P}|}$
\Ensure Average gradient variance $\overline{\mathrm{Var}}$

\State Initialize list $\mathcal{V} \gets \emptyset$ \Comment{store $\mathrm{Var}[\partial_k E]$}

\For{$i = 1$ \textbf{to} $N_s$}
    \State Sample $\boldsymbol{\theta}^{(i)} \in [0,2\pi)^{|\mathcal{P}|}$ uniformly at random
    \State Prepare $\ket{\psi_{\boldsymbol{\theta}^{(i)}}} \gets U(\boldsymbol{\theta}^{(i)})\ket{0}^{\otimes n}$
    \State Compute $E^{(i)} \gets \bra{\psi_{\boldsymbol{\theta}^{(i)}}} H \ket{\psi_{\boldsymbol{\theta}^{(i)}}}$
\EndFor

\For{$k = 1$ \textbf{to} $|\mathcal{P}|$}
    \State Initialize list $\mathcal{G}_k \gets \emptyset$ \Comment{store gradients}
    \For{$i = 1$ \textbf{to} $N_s$}
        \State $\partial_k E^{(i)} \gets \text{Gradient}\!\left(E\right)\big|_{\boldsymbol{\theta}^{(i)}}$
        \State Append $\partial_k E^{(i)}$ to $\mathcal{G}_k$
    \EndFor
    \State $\mathrm{Var}_k \gets \dfrac{1}{N_s}\sum_{i=1}^{N_s}\left(\partial_k E^{(i)}\right)^2$
    \State $\mathcal{V}$.append($\mathrm{Var}_k$) 
\EndFor

\State $\overline{\mathrm{Var}} \gets \dfrac{1}{|\mathcal{P}|}\sum_{v \in \mathcal{V}} v$

\State \Return $\overline{\mathrm{Var}}$
\end{algorithmic}
\end{algorithm}

\iffalse
\begin{itemize}
    \item Sample $N_s$ random parameter vectors $\bm{\theta}, \theta_k\in [0,2\pi),  \forall k$.
    \item Compute the expectation value 
    \begin{equation}
        \bra{\psi_{\bm{\theta}}}E\ket{\psi_{\bm{\theta}}} 
    \end{equation}
    for each $\ket{\psi_{\bm{\theta}}} $.
    \item Use the \emph{Parameter-Shift Rule} to estimate $\partial_kE$ for all $k$'s and for all $N_{s}$ states. This will give as a distribution of the individual derivatives across the whole landscape.
    \item Calculate the average of the squares of the partial derivatives of the cost function in order to get our final estimate for the variance $\text{Var}[\partial_k E]=\braket{(\partial_kE)^2}$ \footnote{Note that for the variance computation we only need to account for the expectation values of the squares of the partial derivatives, since the landscape is unbiased as seen in Eq.\ref{eq: unbiased}} of such distribution.
    \item Average the value of the variance over all trainable parameters.
\end{itemize}
\fi

\subsection{Resource constraints}

The overall resource cost of the ansätze can be decomposed into \emph{classical} (optimization and post-processing time) and \emph{quantum} (circuit width and depth) contributions. In a DSE,  it is crucial to select a well-suited cost function that accounts for all quantities to be minimized (or enforce as constraints) during the optimization \cite{10.5555/2821260}. Accordingly, we summarize these through a single figure that combines three representative cost metrics\footnote{The number of qubits is not considered within the cost metric, since generally in VQAs it is fixed by the problem dimensionality.}, namely: (i) number of trainable parameters, (ii) circuit depth, and (iii) number of two-qubit gates:
\begin{equation}\label{eq: cost_metric}
    \text{Cost} = \alpha \cdot \mathcal{N}(N_{\text{param}})+ \beta\cdot \mathcal{N}(\text{Depth})+ \gamma \cdot \mathcal{N}(N_{2q}).
\end{equation}
Here, $\mathcal{N}(x) = \frac{x-x_{\text{min}}}{x_{\text{max}}-x_{\text{min}}} \in [0,1]$ denotes a min-max normalization computed over the full set of PQCs considered, and $\alpha, \beta, \gamma$ are tunable hyperparameters controlling the relative weight of each component. Unless stated otherwise, we set $\alpha=\beta=\gamma=\frac{1}{3}$, assigning equal priority to all three terms.

The normalized cost in Eq.~\eqref{eq: cost_metric} is used as a resource constraint when identifying Pareto-optimal circuits in the expressibility--trainability plane. In addition, some relevant trade-offs studied in the literature (e.g., trainability--depth) will also be analyzed \cite{McClean2018BarrenPlateaus,Holmes2022ExpressibilityGradients}; in those cases, the different resource metrics of interest will be considered separately and in their unnormalized (absolute) form to extract meaningful insights from them.

%EXPRESSIBILITY-COST RESULTS
\begin{figure*}[t]
    \centering
    \includegraphics[width=1\textwidth]{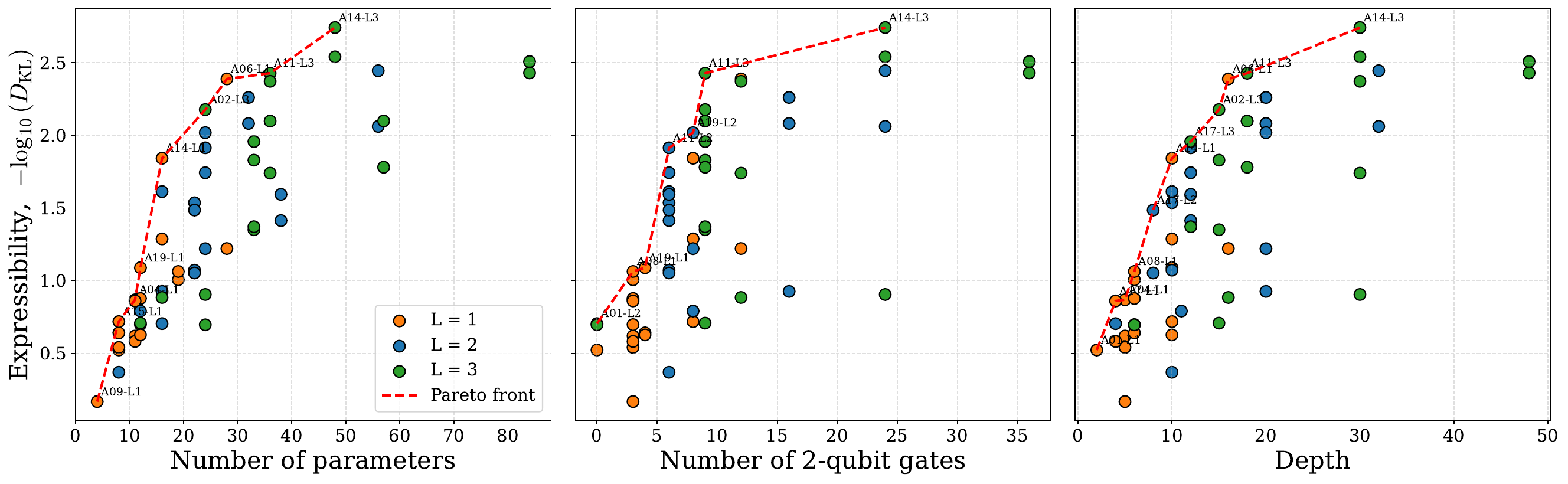}
    \caption{Expressibility vs cost metrics scatter plot for the family of 19 PQCs and number of layers  $L=1,2,3$. The red dashed line connects the pareto-optimal ansätze for each particular trade-off.}
    \label{fig:expr_cost}
\end{figure*}

\subsection{Design Space of Pareto-optimal circuits}\label{sec:DSE}

Once the PQC family has been evaluated, the circuits achieving Pareto-optimality with respect to the aimed metrics are extracted and represented in the design space. In a multi-objective optimization problem, Pareto-optimality is defined via dominance: a solution $S_1$ is Pareto-optimal if there is no solution $S_2$ such that $S_2$ is at least as good as $S_1$ in all objectives and strictly better in at least one. Therefore, any improvement in one metric from a Pareto-optimal circuit necessarily comes at the expense of at least one other metric.

The design space is spanned by \emph{design variables}, i.e., controllable architectural choices that can be readily specified when selecting or constructing an ansatz, while still being representative enough to induce meaningful variations in the objective (metric) space. As illustrated in Figure.~\ref{fig:triangle}, in a DSE, each point in the design space is mapped through a multi-objective function $f:\mathrm{DS}\rightarrow\mathcal{M}$ into a metric space, and it is desirable that nearby designs yield comparable performances (i.e., that the map $f$ exhibits a certain degree of smoothness) \cite{NardiKO2019PracticalDSE}. Moreover, while $f$ is generally not injective (different designs may attain similar metric values), the chosen variables should nevertheless be expressive enough to span (up to a practical boundary of interest) a broad range of attainable objective values. Consistent with established DSE practice, these parameters may be categorical, discrete/integer, or continuous \cite{GautierACK2022Sherlock}. Here, the design space is therefore parameterized by: (i) the primitive (trainable) gate set used to build the ansatz, (ii) the number of layers (i.e., repetitions of the primitive block), and (iii) the required hardware connectivity, represented by an interaction graph that specifies the allowed range of 2-qubit operations.  
\section{Results}

As explained in the prior section, the final goal of this work is to retrieve the subset of Pareto-optimal circuits within our original family accounting for all three metrics concurrently, and then observe their representation in the Design Space. However, it is also crucial to analyze the pairwise trade-offs in order to gain understanding first about the partial interplay between these metrics. 

\subsection{Expressibility-cost}

To begin with, the expressibility values of the PQCs are scattered against different cost parameters in order to see the emerging trends. In every instance of Fig.~\ref{fig:expr_cost}, a clear correlation can be seen, since more expressive circuits generally require more resources, which translates to larger values for all 3 cost metrics. This is consistent with the theory introduced in Sec.~\ref{Sec: II}, since increasing the dimension of the reachable state space of an ansatz comes at the expense of additional trainable parameters in our PQC, leading also to an increasing number of gates and circuit depth. 

\begin{table}[h!]
\centering
\caption{Pareto-optimal circuits for diverse resource-expressibility trade-offs.}
\label{tab:pareto_by_metric}
\setlength{\tabcolsep}{4pt}
\renewcommand{\arraystretch}{1.15}
\begin{tabular}{l p{0.55\linewidth}}
\hline
\textbf{Category} & \textbf{Pareto-optimal Circuit ID's}\\
\hline
Parameters &
A09-L1, A15-L1, A04-L1, A19-L1,\\
& A14-L1, A02-L3, A06-L1, \textbf{A11-L3},\\
& \textbf{A14-L3}\\
\hline
2-qubit gates &
A01-L2, A08-L1, A19-L1, A11-L2,\\
& A19-L2, \textbf{A11-L3}, \textbf{A14-L3}\\
\hline
Depth &
A01-L1, A17-L1, A04-L1, A08-L1,\\
& A17-L2, A14-L1, A17-L3, A02-L3,\\
& A06-L1, \textbf{A11-L3}, \textbf{A14-L3}\\
\hline
\end{tabular}
\end{table}

In addition to this, Pareto-optimal fronts are plotted in Fig.~\ref{fig:expr_cost}, whose elements are further detailed in Table~\ref{tab:pareto_by_metric}. Remarkably, circuits 11 and 14 with $L=3$ are the only ones lying in all three Pareto borders. Interestingly, $A14-L3$ is the most expressive circuit out of the whole set of PQCs considered. Consequently, it lies on the top-right edge of all borders, the most resource-expensive region within the optimal circuits. Moreover, other PQCs that appear (in different layered versions) in at least 2 borders are the 1, 2, 8 and 19, that are characterized by their low number of trainable parameters and two-qubit gates.

Besides Pareto-optimality, there is more information to be read out of Fig.~\ref{fig:expr_cost}. In particular, if two circuits are equally expressive, but one of them presents an overhead in terms of trainable parameters, then it necessarily contains redundant degrees of freedom. Figure~\ref{fig:expr_cost}a reveals this, where the \emph{redundancy} of an ansatz can be quantified as the length of the horizontal line between itself and the Pareto front\footnote{Since a small portion of the PQC space is being considered, even Pareto-optimal circuits could be redundant; hence, the hotizontal distance to the Pareto front constitutes a \emph{lower bound}.}. Such information is crucial since redundant non-optimal models are typically good candidates for pruning techniques \cite{lecun1989optimal, hassibi1993optimal, han2015learning}  akin to those widely used in classical Neural Networks. Such techniques aim to reduce the size of the models while maintaining important properties such as expressiveness.

\begin{figure}[h!]
    \centering
    \includegraphics[width=0.8\linewidth]{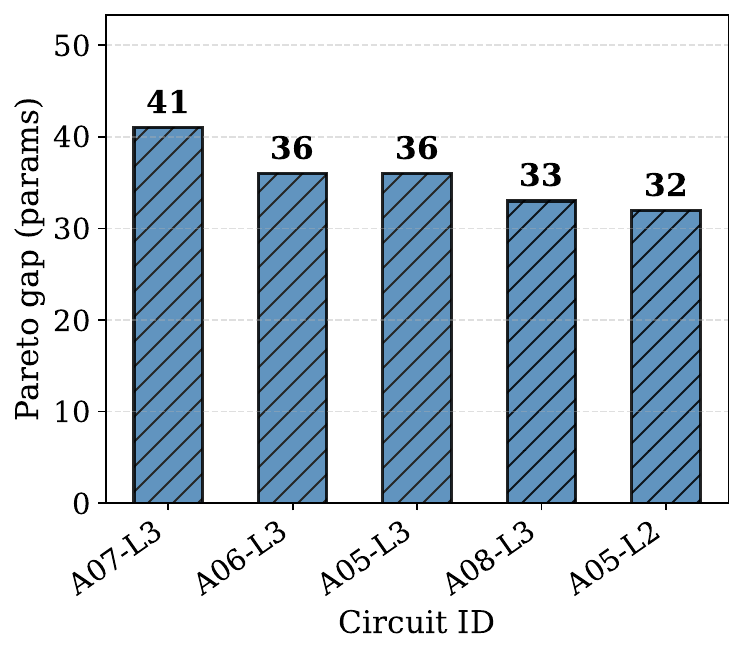}
    \caption{Bar plot showcasing the parameter distance to the Pareto-front of the 5 most redundant circuits.}
    \label{fig:redundant_PQCs}
\end{figure}

Figure~\ref{fig:redundant_PQCs} shows the 5 most redundant circuits, as well as their parameter overhead with respect to the Pareto-optimal ansätze presenting the very same expressive power\footnote{As the expressibility metric is continuous, a linear interpolation of the Pareto-front is carried out in order to find the most accurate projection of the circuits.}. Although 4 different PQCs appear in this ranking, it is important to remark that they can be grouped into 2 categories, as PQCs 7 and 8 present the same structure, only differing by the primitive gate set, and the same occurs for circuits 5 and 6. Furthermore, more layered circuits are found to be more redundant; this is not surprising, as expressibility saturation with the number of layers is a well-known phenomenon \cite{Sim2019Expressibility}.

Up to now, the set of ansätze have been characterized by problem-independent metrics. As introduced at the beginning,  the performance of classical/quantum ML models highly depends on the task where they are being evaluated. It is therefore key to incorporate such missing information for the DSE by considering an additional objective to the optimization: trainability.

\subsection{Trainability-cost}\label{sec: train_depth}

\begin{figure}[h!]
    \centering
    \includegraphics[width=1\linewidth]{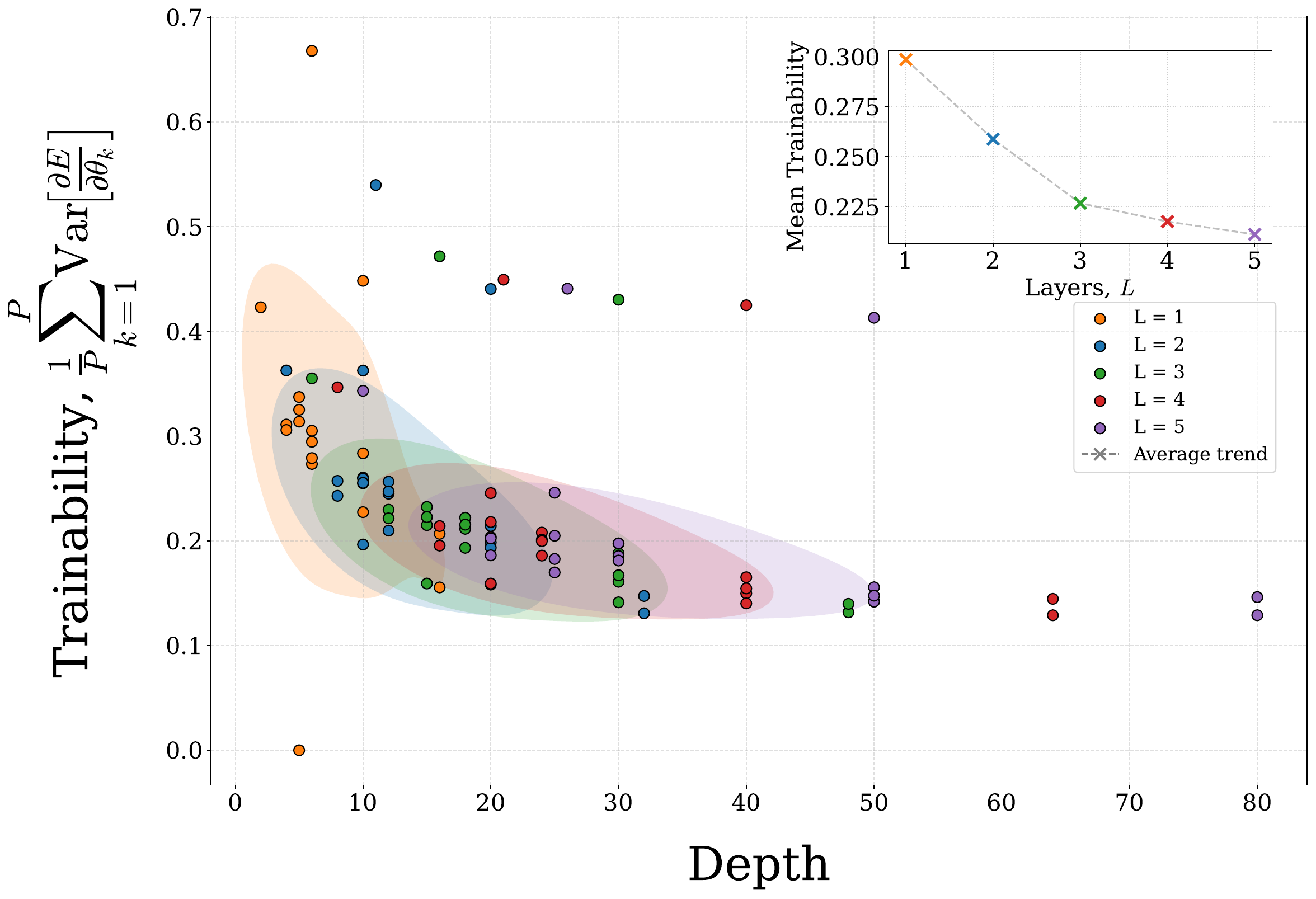}
    \caption{Trainability vs depth scatter plot for the family of 19 PQCs, number of layers  $L\in [1,5]$ and $n=4$ qubits. Colored regions are obtained by 2D Kernel density estimate method with a threshold of $95\%$ probability density. The upper-right plot shows the average trend of trainability values as $L$ increases.}
    \label{fig:trainability_depth}
\end{figure}

As mentioned above, high-dimensional PQCs can present flatter landscapes, leading to worse trainability. Hence, a negative correlation between trainability and cost metrics is expected to be observed. Indeed, this is manifested in Fig.~\ref{fig:trainability_depth}. Because of that, a Pareto border computation is not required, as it is trivial: those circuits with larger trainability (consequently with smaller cost) will be optimal. 

\begin{figure*}[h!]
\centering
\includegraphics[width=0.87\textwidth]{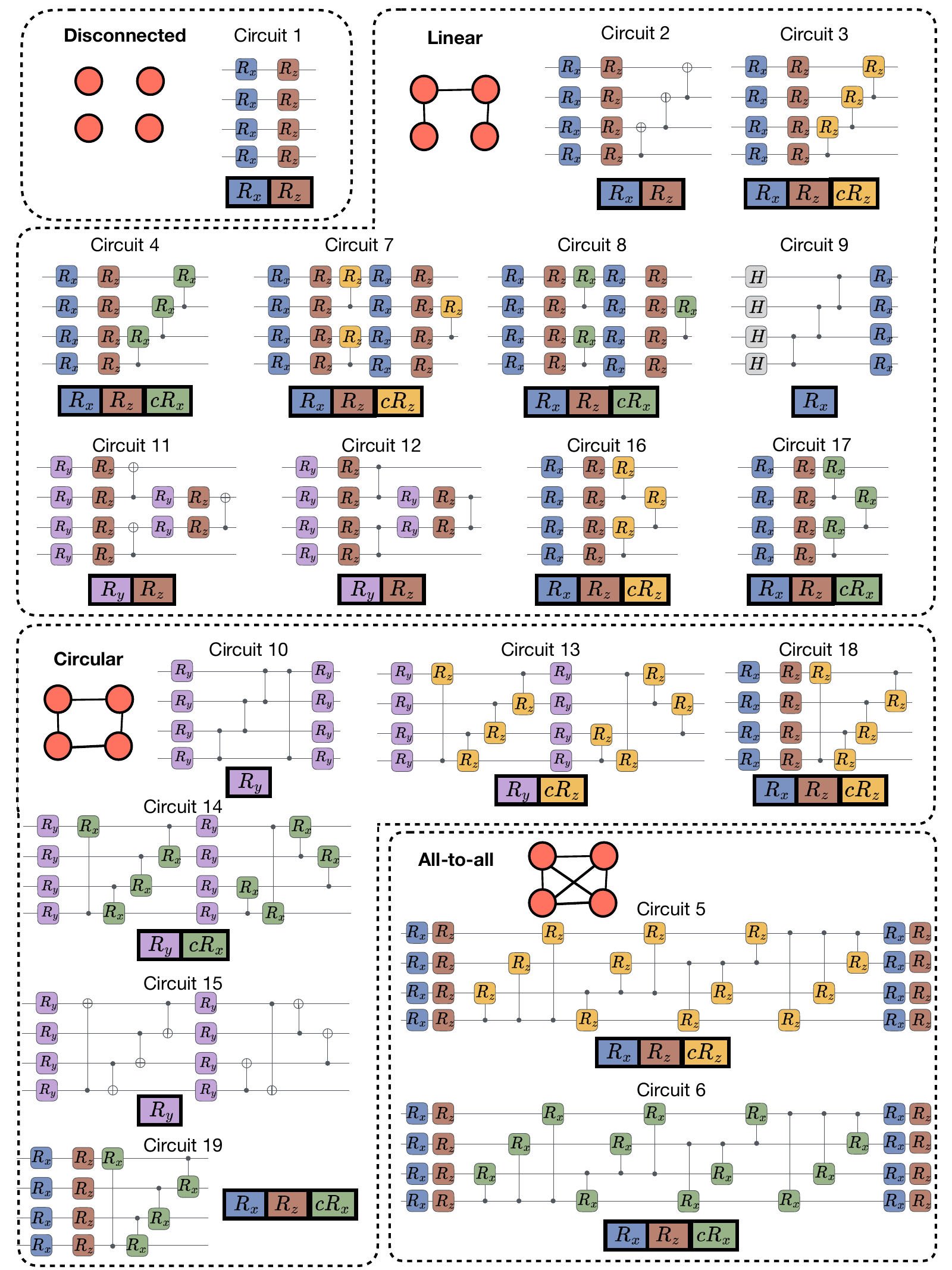}
\caption{Family of parametrized quantum circuits studied in this work and visual representation of their required qubit connectivity and trainable primitive gate set. Set of ansatzes extracted from \cite{Sim2019Expressibility}.}
\label{fig:appendix_circuits}
\end{figure*}

The maximum trainability, quantified by a mean gradient variance of $0.667$, is achieved by ansatz A10 at $L=1$. This finding is consistent across different cost functions; for instance, the same circuit proves optimal for the Heisenberg model with a remarkably similar value of $0.659$ (see Fig.~\ref{fig:unified_analysis}). In the case of the Local-X model, the peak performance shifts slightly to A09-L2, yielding a mean variance of $0.493$. It is important to note that these leading PQCs share a common structure: they employ linear entanglement via non-parametrized CZ gates and restrict all trainable parameters to single-qubit rotations (see Fig.~\ref{fig:appendix_circuits}). This structural simplicity appears to preserve trainability more effectively across different Hamiltonian structures.

Moreover, it can be seen how the per-layer centroids obtained by a 2D kernel density estimation evolve as we increase the number of layers. Such trends mark a clear inverse relationship in which increasing $L$ on average displaces the circuits into a low-trainability region within the metric space until a final saturation value. In particular, if the circuit in the high-$L$ regime is approximately a 2-design, such saturation value is known to decrease exponentially with the number of qubits \cite{McClean2018BarrenPlateaus}. 

However, not all PQCs converge to the same trainability value as the number of layers increases. Table~\ref{tab:top5_trainability_ising_L5} reports the circuits whose $L=5$ versions yield the largest mean gradient-variance values, providing insight into which PQCs exhibit the best trainability saturation. Notably, ansätze 10 and 15 achieve the highest values of the metric (0.440 and 0.413), corresponding to increases of 108.5$\%$ and 95.7$\%$ over the average across all circuits (0.211), respectively (i.e., about $2.09\times$ and $1.96\times$ the mean). In addition, interestingly, the top four circuits (the only ones whose trainability lies above the overall average) are precisely the ansätze with the \textit{smallest} number of trainable parameters. Also, none of them includes parametrized two-qubit gates (See Fig.\ref{fig:appendix_circuits}). This particular feature naturally constrains expressibility, while simultaneously helping the circuit avoid approaching the 2-design limit at large $L$, thereby preventing trainability degradation in deep regimes \cite{Cerezo_2021}.

\begin{table}[h!]
\centering
\caption{Most trainable circuits at $L=5$.}
\label{tab:top5_trainability_ising_L5}
\setlength{\tabcolsep}{6pt}
\renewcommand{\arraystretch}{1.15}
\begin{tabular}{l c c}
\hline
\textbf{Circuit ID} & \textbf{Trainability} & \textbf{$\#$Parameters}\\
\hline
A10-L5 & 0.440& 40\\
A15-L5 & 0.413& 40\\
A01-L5 & 0.343& 40\\
A09-L5 & 0.246& 20\\
A03-L5 & 0.205& 44\\
\hline
Mean-L5  & 0.211 & --\\
\hline
\end{tabular}
\end{table}

On the other hand, at the deep-circuit limit ($L=5$), the mean saturation values for the gradient variance under the Heisenberg and Local-X models are $0.257$ and $0.129$, respectively. The observation that the Local-X model—composed strictly of local terms—exhibits lower saturation points than the more complex Heisenberg model is somewhat counter-intuitive. However, it is important to consider that the relation of cost locality on trainability is linked to the qubit scaling  $n$ \cite{Cerezo_2021}. Consequently, while the Local-X model shows suppressed trainability in this $n=4$ regime, these trends may change as the system size increases toward the thermodynamic limit.

\subsection{Expressibility-trainability and Pareto-optimal architectural Design Space}\label{sec: expr_trainability}

\begin{figure*}[h!]
    \centering
    \includegraphics[width=1\textwidth]{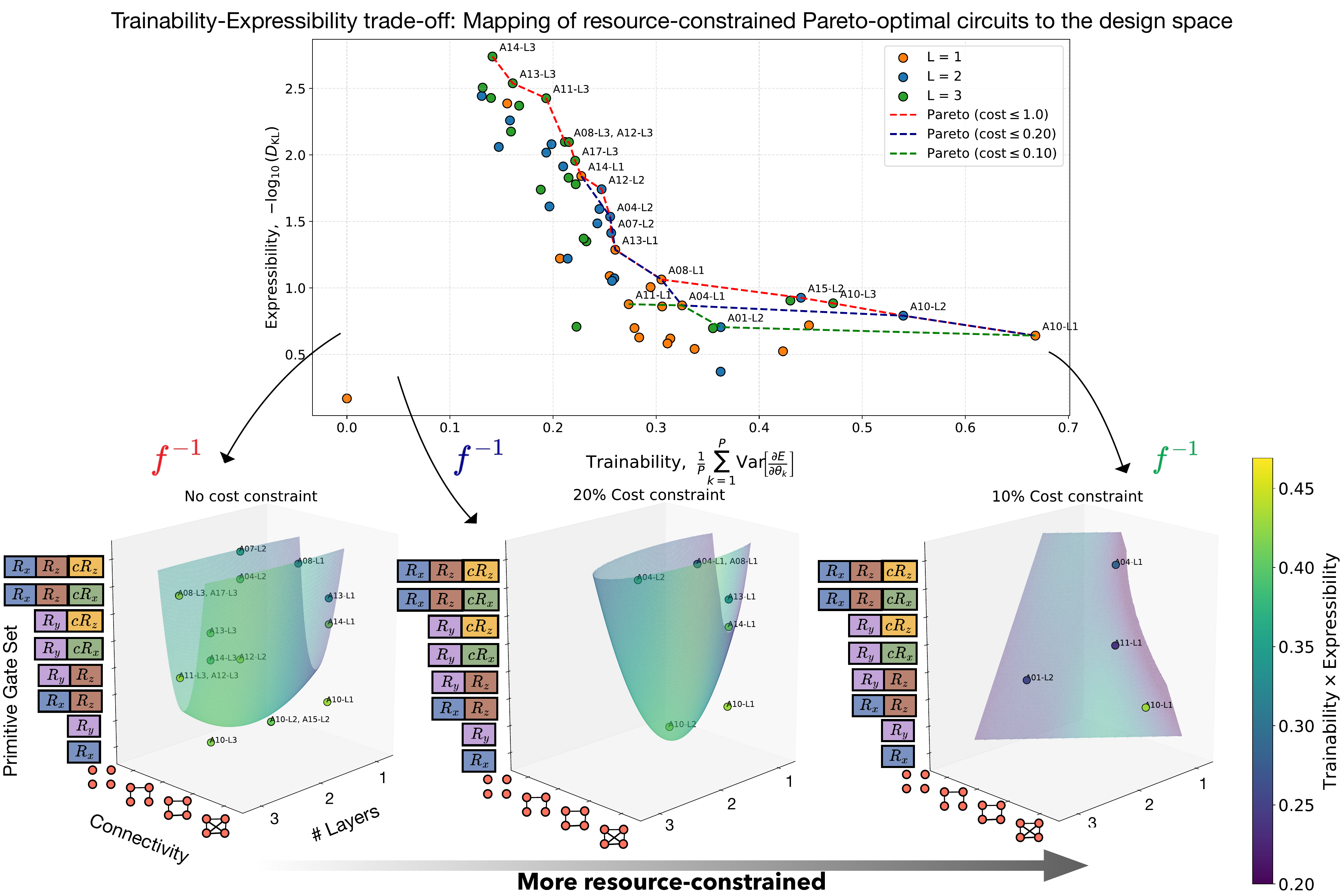}
    \caption{Expressibility--trainability scatter plot for a family of 19 parameterized quantum circuits (PQCs) at depths $L=1,2,3$. The red, blue, and green dashed segments connect the Pareto-optimal ansätze computed under different resource constraints. These ansätze are embedded in the corresponding 3D design space (number of layers, connectivity pattern, and primitive gate set), where points are colored by the composite score $\mathrm{Trainability}\times \mathrm{Expressibility}$. A polynomial surface $z=g(x,y)$ is fitted to the mapped Pareto-optimal points (total degree $2$ for the first two constraints and degree $1$ for the most restrictive case). The surface colormap is predicted by an MLP regressor with two hidden layers of $32$ units each and $\tanh$ activation, trained on the full set of PQC samples (after standardizing the inputs).}
    \label{fig:expr_trainability}
\end{figure*}

Finally, a global view of the problem is considered. Figure~\ref{fig:expr_trainability} displays all ansätze in the expressibility--trainability plane, from which Pareto-optimal circuits are identified. This analysis is carried out under three resource settings: unconstrained, and constrained to circuits whose cost does not exceed either $20\%$ or $10\%$ of the maximum (red, blue and green dashed segments, respectively). Here, the maximum cost (normalized to $\mathrm{Cost}=1$) corresponds to ansätze 5 and 6 at $L=3$, which contain 36 two-qubit gates, have depth 48, and feature 84 trainable parameters.

Table~\ref{tab:pareto_by_cost_constraint} reports the resulting Pareto-optimal sets pertaining to each front in Figure~\ref{fig:expr_trainability}. As expected, the number of PQCs decreases as the cost budget is tightened.  Indeed, for the case $\text{Cost}\leq 10\%$, the Pareto set reduces to only four circuits (A10-L1, A04-L1, A01-L2, A11-L1), reflecting that under severe resource constraints the optimal designs are forced toward low-depth, low-parameter structures (see Fig.~\ref{fig:appendix_circuits}). In the unconstrained case, the Pareto front spans a wide range of trade-offs, from highly expressive but moderately trainable circuits (upper-left region) to highly trainable but weakly expressive ones (right-lower region). In particular, the best composite scores $\mathrm{Expr}\times\mathrm{Train}$ are achieved by A11-L3, A12-L3, A08-L3, and A17-L3 (0.469, 0.452, 0.444, and 0.433, respectively). Additionally, the top 3 circuits with highest score are the same for the Heisenberg model in the exact same order (with scores of 0.621, 0.575 and 0.562). Interestingly, they all share the number of layers $L=3$ and circuit connectivity (linear), which resembles de qubit interaction within both TFIM and Heisenberg Hamiltonians. On the other hand, the overall scores for the Local-X model are in general lower (as expected from the conclusions discussed in Sec.\ref{sec: train_depth}), where the largest value corresponds to A08-L3 with $\mathrm{Expr}\times\mathrm{Train} = 0.271$. Due to its difference in structure, a different PQC appear within the top 5 best scores; A10-L3 (with no parametrized two-qubit gates and a circular ciurcuit connectivity) with $\mathrm{Expr}\times\mathrm{Train} = 0.246$.

\begin{table}[h!]
\centering
\caption{Pareto-optimal circuits under different normalized cost constraints.}
\label{tab:pareto_by_cost_constraint}
\setlength{\tabcolsep}{4pt}
\renewcommand{\arraystretch}{1.15}
\begin{tabular}{l p{0.55\linewidth}}
\hline
\textbf{Resource constraint} & \textbf{Pareto-optimal Circuit ID's}\\
\hline
Unconstrained &
A11-L3, A12-L3, A08-L3, A17-L3, A12-L2, A10-L1, A10-L2, A14-L1,\\
& A10-L3, A13-L3, A15-L2, A04-L2, A14-L3, A07-L2, A13-L1, A08-L1\\
\hline
$\leq 0.20$ &
A10-L1, A10-L2, A14-L1, A04-L2, A13-L1, A08-L1, A04-L1\\
\hline
$\leq 0.10$ &
A10-L1, A04-L1, A01-L2, A11-L1\\
\hline
\end{tabular}
\end{table}

It is also worth noting that the only circuit that appears in all Pareto fronts  in Fig.~\ref{fig:expr_trainability} is A10-L1. That is, it presents a cost of under $10\%$ (consisting of only 4 two-qubit gates, 8 trainable parameters and a depth of 6) of the maximal while satisfying Pareto-optimality, with a score of $\mathrm{Expr}\times\mathrm{Train} = 0.428$, top 6 within the entire set. Analogously, A10-L2, A14-L1, A04-L2, A13-L1, and A08-L1 are shared between the unconstrained and the $20\%$-budget fronts.

Finally, the Pareto-optimal families are mapped back into the design space (bottom panels of Fig.~\ref{fig:expr_trainability}) to localize the regions that yield the best expressibility-trainability trade-offs. As the cost constraint becomes tighter (left to right), the optimal points progressively concentrate toward shallower designs with simpler gate sets and reduced connectivity requirements. In the most restrictive case ($\mathrm{Cost}\leq 10\%$), the Pareto set collapses to circuits that are either single-layered (A04-L1, A11-L1, A10-L1) or exhibit sparse interaction graphs with linear/disconnected connectivity (A01-L2, A04-L1, A11-L1).

To provide a complete description of these optimal regions and enable extrapolation beyond the evaluated templates, we fit, for each constraint scenario, a low-degree polynomial surface to the Pareto-optimal points in design space (Fig.~\ref{fig:expr_trainability}). Then, an estimation of the score $\mathrm{Expr}\times\mathrm{Train}$ is performed over the different surfaces by learning a regression model on the entire family of $19\times 3=57$ PQC instances and predicting scores for continuous set of designs across the surface. The resulting Pareto surfaces offer an interpretable approximation of the optimal architectural manifold: high-score regions (yellow) represent areas where novel circuit designs are expected to achieve favorable trade-offs, thus providing an actionable guideline for subsequent ansatz synthesis.

\subsection{Exploring the use for optimal ansatz synthesis}

\begin{figure*}[h!]
\centering
\includegraphics[width=0.75\textwidth]{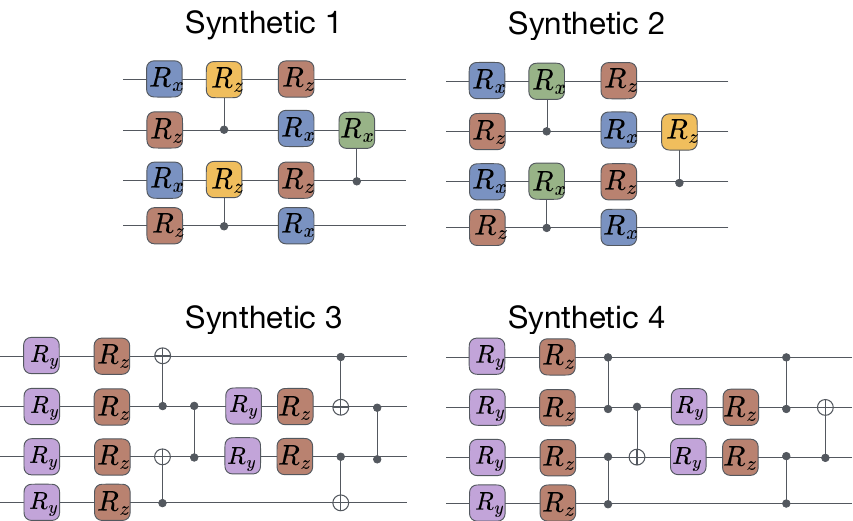}
\caption{Ilustration of the proposed synthetic circuits.}
\label{fig:synthetic_PQCs}
\end{figure*}

Based on the results discussed in Sec. \ref{sec: expr_trainability}, the circuits achieving the highest scores were identified to be A11-L3, A12-L3, A08-L3, and A17-L3. To explore the extended use of the presented methodology to guide and inform PQC synthesis, we perform a targeted synthesis of new candidate circuits. By introducing slight structural variations to these high-performing PQCs while maintaining proximity within the design space preserving $L=3$ layers, linear connectivity, and specific primitive gate sets (as suggested by the surfaces in Fig. \ref{fig:expr_trainability}), four synthetic circuits, denoted as S01-S04, are proposed (see Fig. \ref{fig:synthetic_PQCs}).

\begin{figure}[h!]
    \centering
    \includegraphics[width=1\linewidth]{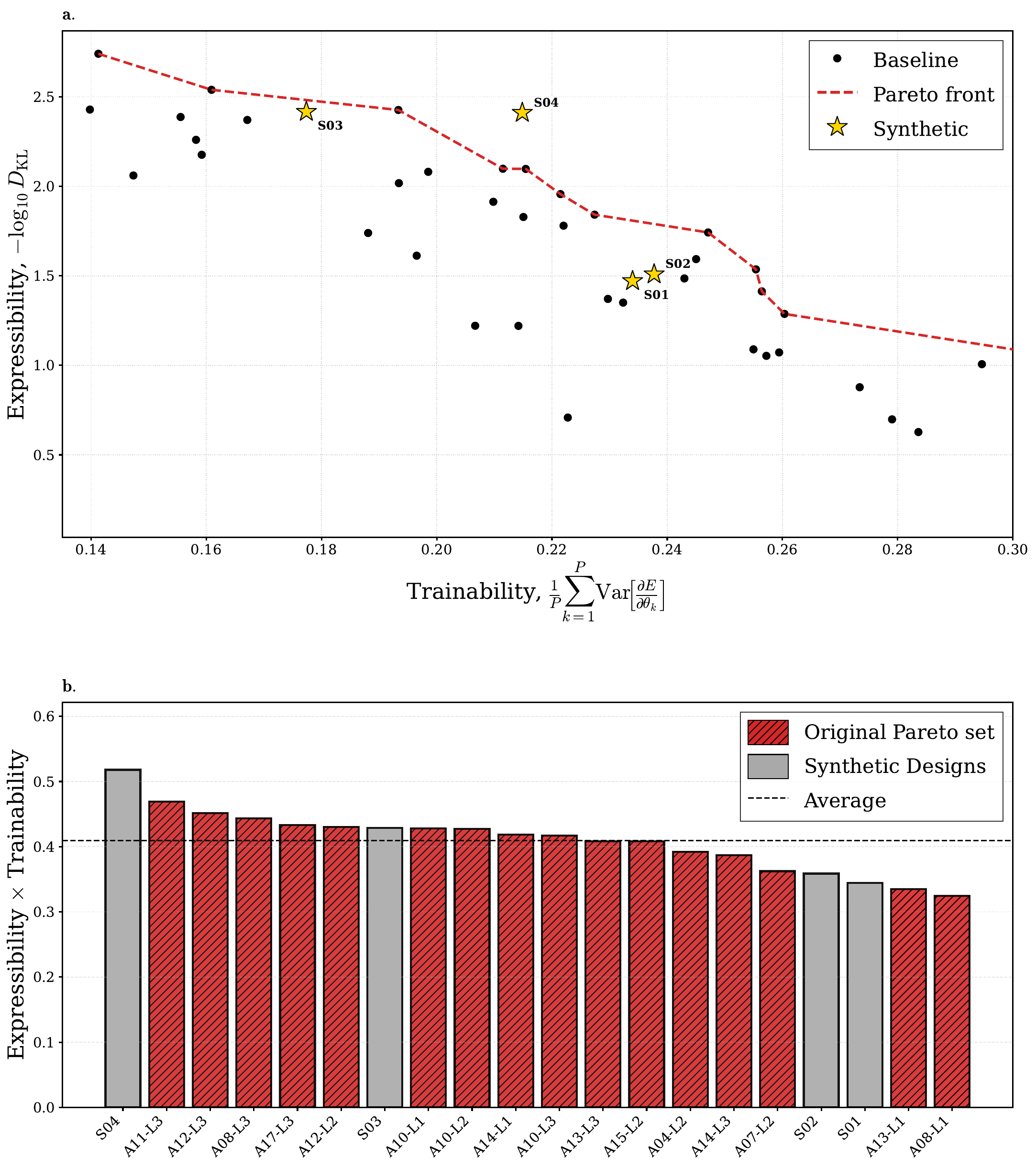}
    \caption{Comparative performance analysis of synthetic and baseline circuits. \textbf{a.} Distribution of quantum circuits in the expressibility--trainability landscape. The baseline PQCs (black) are compared against synthetic designs (yellow stars), where the dashed line denotes the Pareto front of the initial search space. \textbf{b.} Performance ranking based on the metric product ($\text{Score} = \text{Expressibility} \times \text{Trainability}$), comparing the Pareto-optimal baseline ansätze against the proposed synthetic PQCs.}
    \label{fig:synthetic}
\end{figure}

The performance of these synthesized designs is evaluated and compared against the original search space in Fig. \ref{fig:synthetic}. Notably, in the expressibility--trainability plane (Fig. \ref{fig:synthetic}a), the synthetic circuit S04 surpasses the initial Pareto front, establishing itself as a new Pareto-optimal ansatz. Furthermore, the ranking in Fig. \ref{fig:synthetic}b shows that all four synthetic designs achieve scores higher than the minimum within the original Pareto-optimal set. Specifically, S03 performs above the population average, and S04 achieves the top-ranked position, outperforming the best original ansatz by $10.42\%$.

This brief example illustrates that the proposed DSE framework can be effectively employed to synthesize new optimal ansätze. While this study explored a limited set of variations, a more rigorous and systematic application of this synthesis protocol is expected to yield even more optimal structures tailored to specific hardware and task constraints.

\section{Discussion and future work}

This work reframes ansatz selection for variational quantum algorithms as an explicit ansatz-design problem: rather than treating circuit choice as a heuristic decision, we evaluate a diverse set of PQCs jointly through expressibility, trainability, and resource cost, and select the outcome via Pareto optimality. The resulting fronts make the well-known tension between ansatz expressiveness and gradient magnitudes quantitatively visible. Crucially, the elements of such fronts are mapped into the design space, where one can identify regions in which circuits offer consistently favorable trade-offs. In this sense, the DSE perspective helps narrowing the attention to families of PQCs that are simultaneously plausible to optimize and realistic to implement while maintaining expressiveness. The practical utility of identifying optimal regions in the design space is empirically illustrated with a toy example: by synthesizing four circuits within Pareto-favorable regions, we already obtain PQCs that achieve improved trade-offs relative to the original set. This result provides an initial validation of the approach as a constructive design tool, and motivates a more systematic and rigorous search for task-oriented ansatzes that jointly optimize trainability, expressibility, and circuit cost while remaining compatible with both near-term and future hardware constraints.

A natural limitation of the proposed methodology is the scalability of the numerical estimate of expressibility and trainability, due to the increasing requirement in the samples \cite{Sim2019Expressibility}. However, recent methods point to viable routes around this bottleneck, including data-driven estimation of expressibility \cite{GNN_expressibility} (exploiting Graph Neural Networks) and efficient trainability evaluation \cite{efficient_trainability} that leverages classical simulability in structured regimes (e.g., Clifford-based circuits). These advances suggest that the DSE methodology can be scalable to larger systems without sacrificing statistical reliability.
\section*{Acknowledgments}
This work has been financially supported by the Ministry for Digital Transformation and of Civil Service of the Spanish Government through the QUANTUM ENIA project call - Quantum Spain project, and by the European Union through the Recovery, Transformation and Resilience Plan - NextGenerationEU within the framework of the Digital Spain 2026 Agenda. Also, by the Generalitat Valenciana grant CIPROM/2022/66 and by the Spanish Ministry of Science, Innovation and Universities and European ERDF under grant PID2024-158682OB-C31. In addition, authors acknowledge funding from the EC through HORIZON-EIC-2022-PATHFINDEROPEN-01-101099697 (QUADRATURE), and the ICREA Academia Award 2024-2028 from the Catalan Government. 
\bibliographystyle{ieeetr}
\bibliography{bib/main}

\clearpage
\onecolumn
\appendices

\section{Set of parametrized quantum circuits}

\newpage
\section{Alternative Hamiltonians simulation}
\label{apx:other_models}

To demonstrate the robustness and generalizability of the proposed framework, we evaluate the expressibility and trainability of the PQCs under different cost functions. This allows one to observe how the previously identified trade-offs behave when subjected to Hamiltonians with distinct physical structures.

First, the isotropic \textbf{Heisenberg model} is considered, defined as:
\begin{equation}
    H_{\text{Heis}} = \sum_{i=1}^{n-1} \left( \sigma_i^x \sigma_{i+1}^x + \sigma_i^y \sigma_{i+1}^y + \sigma_i^z \sigma_{i+1}^z \right) \, .
\end{equation}
In contrast to models dominated by a single interaction type (such as the $\sigma^z \sigma^z$ terms in the Ising Hamiltonian), the Heisenberg model treats all spatial dimensions equivalently. This prevents the optimization process from developing a bias toward specific rotation angles or basis states, providing an objective measure of the ansatz's performance. Given its fundamental relevance in condensed matter physics and interaction profile, this model serves as a representative benchmark for VQE, as demonstrated in foundational studies such as \cite{Nakaji2021expressibilityof}.

Secondly, a \textbf{Local-X model} is also analyzed. This model is defined as the limit of a Transverse Field Ising Model (TFIM) where the nearest-neighbor interaction term is suppressed ($J=0$), resulting in a sum of non-interacting transverse fields:
\begin{equation}
    H_{\text{Local-}X} = \sum_{i=1}^{n} \sigma_i^x \, .
\end{equation}
The selection of this model is motivated by the theoretical results in \cite{Cerezo_2021}, which suggest that the appearance of Barren Plateaus can depend on the locality of the cost function.

\begin{figure*}[h!]
    \centering
    \includegraphics[width=0.9\textwidth]{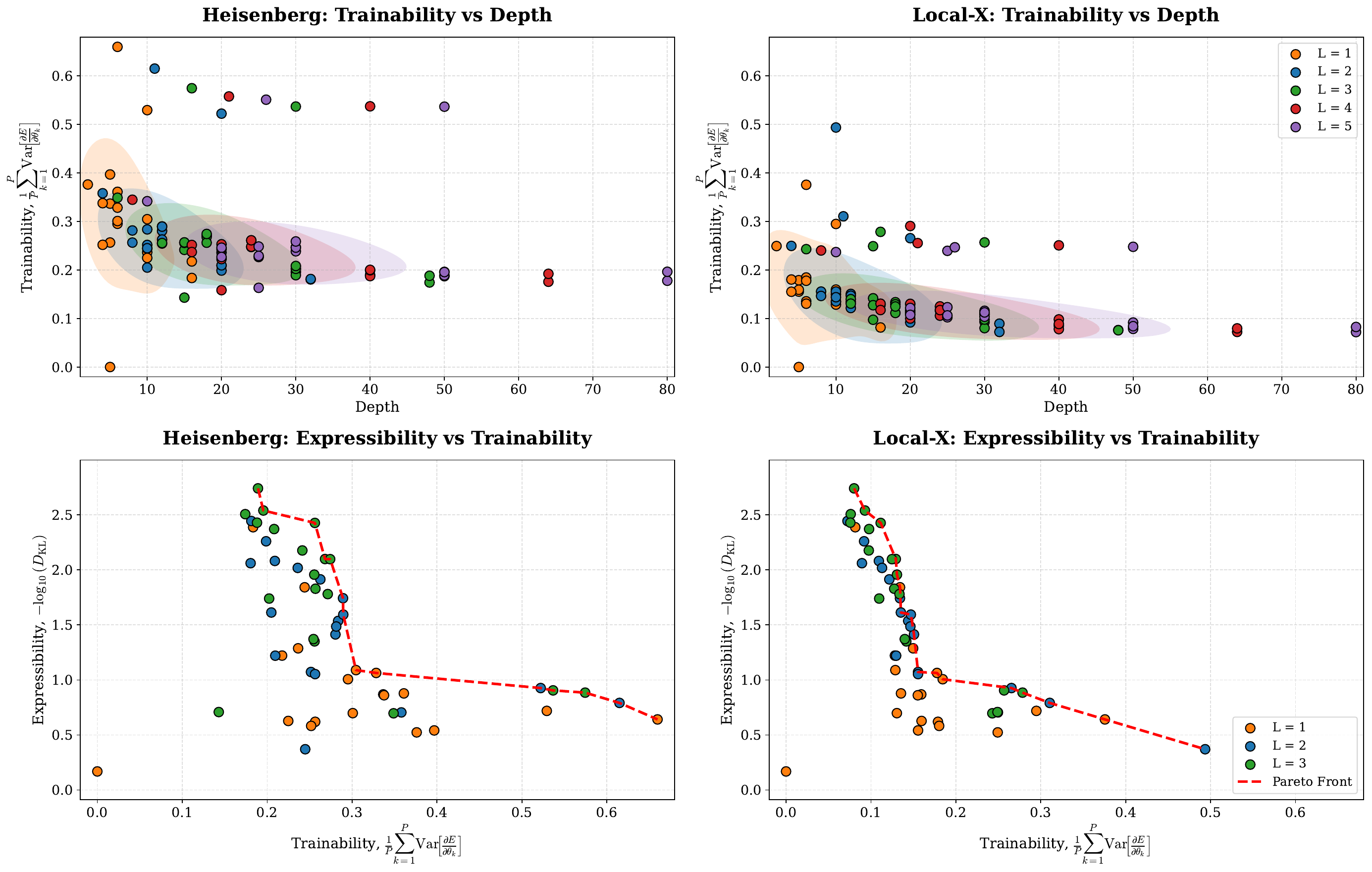}
    \caption{\textbf{Comprehensive analysis of the Heisenberg and Local-X models.} 
    The top row illustrates the relationship between circuit depth and trainability, quantified by the mean gradient variance $ \frac{1}{P}\sum_{k=1}^{P} \mathrm{Var}[\partial \theta_k E] $. Shaded regions represent the Kernel Density Estimation (KDE) at a 95\% density threshold for each layer $L \in \{1, \dots, 5\}$. 
    The bottom row shows the trade-off between expressibility ($-\log_{10} D_{\text{KL}}$) and trainability for $L \in \{1, 2, 3\}$. The dashed red line indicates the Pareto front, identifying ansatzes that maximize both metrics simultaneously.}
    \label{fig:unified_analysis}
\end{figure*}

Finally, it is important to note that the following results do not include the \textit{expressibility--cost} trade-off, as it solely depends on the PQCs structure and does not vary with the cost function. Hence, only the \textit{trainability--cost} and \textit{expressibility--trainability} relationships will be considered. By evaluating these specific dependencies, as illustrated in Fig. \ref{fig:unified_analysis}, one can effectively benchmark how different circuit structures adapt to varying landscape complexities, providing a comprehensive guide for selecting optimal PQCs in practical VQA implementations.

\section{Examples of Synthesized PQCs}

\end{document}